\documentstyle[epsfig]{aipproc}

\begin{document}
\title{Status of the CANGAROO-III Project}

\author{Masaki Mori$^*$\footnote{%
E-mail: {\tt morim@icrr.u-tokyo.ac.jp}}
 for the CANGAROO Collaboration\footnote{%
Ibaraki University; Ibaraki Prefectural University of Health
Sciences; Institute
of Space and Astronautical Science; Osaka City University;
Kyoto University; Konan University; National Astronomical Observatory
of Japan; Tokai University; Tokyo Institute of Technology;
Insititute for Cosmic Ray Research, University of Tokyo;
STE laboratory, Nagoya University; Yamagata Universiry;
Yamanashi Gakuin University;
Institute of Physical and Chemical Research; University
of Adelaide; Australian National University
}}
\address{$^*$Institute for Cosmic Ray Research, University of Tokyo,\\
5-1-5 Kashiwanoha, Kashiwa, Chiba 277-8592, Japan}

%\lefthead{LEFT head}
%\righthead{RIGHT head}
\maketitle

\begin{abstract}
The CANGAROO-III project, a system of four 10~m telecopes
dedicated for gamma-ray astrophysics,
has started in 1999 and is expected to complete in 2004
in Woomera, South Australia.
We report the construction work during the first year,
which includes the completion of the first 10~m telescope 
built as a 7~m telescope in 1999,
and the work in progess to increase the performance by
constructing three more similar telescopes based on the experience
gained during the construction of the first telescope.
\end{abstract}

\section*{Introduction}

The CANGAROO-III is a project to study celestial gamma-rays
in the 100 GeV region utilizing a stereoscopic observation
of Cherenkov light flashes with an array of four 10-meter
telescopes \cite{Mori99a}, following the CANGAROO-I (3.8~m) \cite{Hara93} and
CANGAROO-II (7~m) telescopes \cite{Tanimori99} \cite{Mori99} \cite{Kubo99}
in Woomera, South Australia ($136^\circ47'$E, $31^\circ06'$S,
 160m a.s.l.)\footnote{Visit our website for the latest information:
{\tt http://icrhp9.icrr.u-tokyo.ac.jp}}.

\begin{figure}[t!] % fig 1
\centerline{\epsfig{file=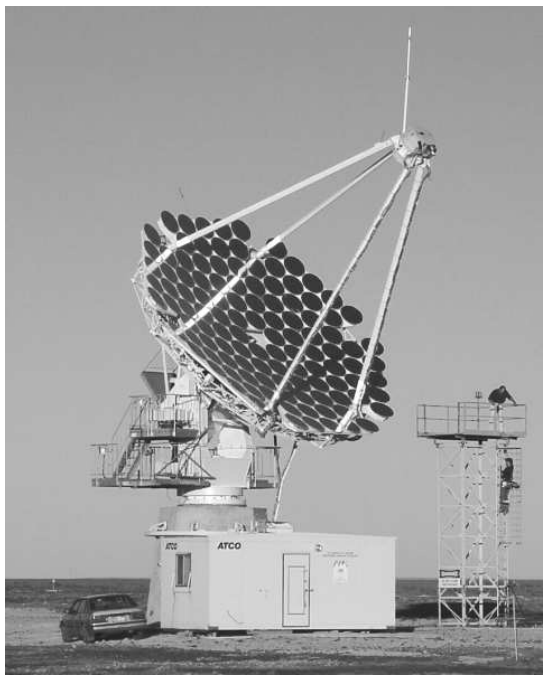,height=3.5in}}
\vspace{10pt}
\caption{The 10~m telescope completed in February 2000 in Woomera,
South Australia. The reflector consists of 114 spherical
mirrors of 80~cm radius. The imaging camera is set at the focus
($f=8$~m) and can be accessed from the tower (right) for maintainance.}
\label{fig:10m}
\end{figure}

It has officially started in April 1999 and is planned as a five-year
program.  In February 2000 we have expanded the
7~m telescope to 10~m, which is the first telescope of the
CANGAROO-III array (Fig.\ \ref{fig:10m}). This year we are building the
second telescope in Japan which will be shipped and installed in 2001.
The other two telescopes will be installed in the fourth and fifth
years. 
Each telescope will be set on a corner of a diamond of about 100~m side
in order to have a maximum number of pairs of telescopes of the
same baseline length.
The first stereoscopic observation will be performed in 2002 and the full
four telescope will be in operation in 2004.

\section*{The First 10~m Telescope}

\noindent{\large\bf 10~m reflector}

The reflector of the 7~m telescope is described elsewhere \cite{Kawachi99}:
its support structure was designed to be
10~m in diameter with an 8~m focal length
and the extension was simple: just adding 54 mirrors of
80~cm in diameter and increasing the counterweights.
However, mounting mirrors on the reflector without dismantling
the telescope was not an easy task since we had to work on
a high platform.
We completed this work in two weeks using a large crane.

Adjusting the mirrors to a common focus was also a challenge.
A flood light placed 5~km away worked as 
a light source providing parallel light.
Instead of covering all mirrors except the mirror being worked on
which was the method used in tuning the 7~m telescope,
we adjusted mirrors using a mask to select a image of 
each mirror 40cm before the focal plane.
Then we could tune the attitude of each mirror by stepping-motor control,
so that the image appeared at the expected position on the mask .
The image of a star was measured with a CCD camera
and its size (FWHM) is $0.20^\circ$. This is a little worse
than the 7~m telescope but expected because of the elongated images
formed by the outer mirrors.

\medskip
\noindent{\large\bf Camera}

We have added 40 PMTs to the prime focus camera to enlarge the
field of view of the camera. Now the camera has 552 PMTs of 
half-inch diameter and
subtends about 3 degrees in octagon shape.
The camera housing itself was
replaced in order to reject stray light coming from
outside of the reflector.

\medskip
\noindent{\large\bf Electronics}

Now the signals from PMTs are fed into analog-buffer amplifiers.
One output goes to the existing front-end module (discriminator and 
scaler) and the other goes to newly developed VME-based ADCs 
(9U height, 32~ch/board, 12~bit resolution, 0.25~pC/count, 
150~ns internal delay, 50~ns gate width), thus enhancing 
the pulse height resolution and dynamic range.
The discriminated signals are sent to TDCs to measure timing
and pulse width as before.

\medskip
\noindent{\large\bf Observation}

Observation resumed in March 2000 with the 10~m reflector
and ADCs are working since April 2000.
Target objects were selected from our list of
TeV gamma-ray souces: SN1006, PSR1706-44 and RXJ1713.7-3946
in order to confirm our previous detections with
the 3.8m telescope (see Ref.\cite{Mori99a} for a list of 
observations with the 3.8m telescope).
Also nearby X-ray selected BL Lacs: PKS2005-489, PKS2155-304 
were observed along with multiwavelength campaigns.
Data analysis is underway and will be reported soon.
A preliminary comparison of data with a Monte Carlo
simulation suggests the threshold energy for a proton
shower is around 600 GeV.

\section*{Work in Progress for New Telescopes}

\noindent{\large\bf Mirrors}

The CFRP mirrors of 80~cm in diameter and $16\sim17$~m in 
curvature radius developed for the CANGAROO-II
telescope are very light and durable \cite{Kawachi99}.
Refining the manufacturing process of them are
underway so as to produce better quality mirrors
and to obtain a better yield.

\medskip
\noindent{\large\bf Camera design}

The new design of a camera will be in hexagonal shape so that
the dead space between PMTs are minimal. The field-of-view
will be $4.3^\circ$ with 427 3/4" PMTs (Hamamatsu R3478)
of size larger than those of
the CANGAROO-II telescope. 
Light guides to increase the light collection are redesigned
to match the new PMT arrangement.
Signals from PMTs are amplified
at the camera and sent to the electronics via twisted-pair
cables to minimize the weight. High voltages are supplied to
PMTs individually from the ground electronics
so that the gains can be tuned remotely.
Now we will apply positive high voltages to PMTs to
solve the discharging problem when the humidity is high
in winter.

\medskip
\noindent{\large\bf Electronics}

The new electronics system will be all VME-based.
The front-end circuit (under development, 16~ch/VME-9U) 
amplifies the signal and feeds
to an ADC (the improved type of those used in the CANGAROO-II 
with faster conversion), 
discriminates it and feeds to a TDC (64~ch/VME-6U), an internal
scaler and a trigger circuit.
The new pattern trigger circuit using PLD devices
are under development to reduce unwanted triggers caused
by night sky background.

Increased data size requires faster data acquisition.
Now we are testing several possibilities including
VME-based Pentium CPU board running a Linux operating system,
which shows faster task switching, reducing deadtime
of data acquisition.

\medskip
\noindent{\large\bf Monitor}

Event data will be sampled and displayed on-line at a
central control room. Scaler data
which can act as a star sensor are acquired independently, 
reducing the load on the data-taking computers.
A blue LED set at the pole of the reflector is used for
field-flattening of the camera sensitivity every night.
A cloud monitor which detects far infrared light \cite{Dowden97} 
will be used to monitor sky condition. 
Weather data and temperatures in
electronics will be recorded as house-keeping data.
These environmental monitor data are acquired through network and stored
with event data.
Telescope tracking will be controlled by networked computers
and its accuracy will be checked with optical CCD cameras by
observing stars.
A light pulser source to monitor telescope
performance has been developed and is being tested 
at the 10m telescope \cite{Patterson00}. 

\section*{Summary}

The CANGAROO-III will be the first array of telescopes in the
southern hemisphere exploring sub-TeV region of the sky.
The HESS array in Namibia, also under construction,
is very welcome since the difference
in time zone makes the sky coverage for transient sources wider,
and should provide useful confirmation.

\end{document}